\documentclass[12pt]{article}
\usepackage{amsmath}
\usepackage{times}
\usepackage{graphicx}
\usepackage{color}
\usepackage{multirow}
\usepackage[authoryear]{natbib}
\usepackage{rotating}
\usepackage{bbm}
\usepackage{latexsym}
\usepackage{fmtcount} 
\newcommand{\citealtt}[1]{\citeauthor{#1},\citeyear{#1}}

\newcommand{\be}{\begin{eqnarray}}
\newcommand{\ee}{\end{eqnarray}}
\renewcommand{\bar}{\overline}
\newcommand{\captionfonts}{\normalsize}

\makeatletter  
\long\def\@makecaption#1#2{%
  \vskip\abovecaptionskip
  \sbox\@tempboxa{{\captionfonts #1: #2}}%
  \ifdim \wd\@tempboxa >\hsize
    {\captionfonts #1: #2\par}
  \else
    \hbox to\hsize{\hfil\box\@tempboxa\hfil}%
  \fi
  \vskip\belowcaptionskip}
\makeatother   

\long\def\symbolfootnote[#1]#2{\begingroup%
\def\thefootnote{\fnsymbol{footnote}}\footnote[#1]{#2}\endgroup}
\renewcommand{\thefootnote}{\normalsize \arabic{footnote}} 	
\begin{document}
\hspace{13.9cm}1

{\LARGE \center The evolution of representation in simple cognitive networks}

\ \\
{\bf \large Lars Marstaller$^{\displaystyle 1,\star}$, Arend Hintze$^{\displaystyle 2, \displaystyle 3, \displaystyle 4,\star}$ \& Christoph Adami$^{ \displaystyle 2, \displaystyle 3}$ 
}\\
{$^{\displaystyle 1}$Department of Cognitive Science, Macquarie University, Sydney, Australia}\\
{$^{\displaystyle 2}$Microbiology \& Molecular Genetics, Michigan State University, East Lansing, MI}\\
{$^{\displaystyle 3}$BEACON Center for the Study of Evolution in Action, Michigan State University, East Lansing, MI}\\
{$^{\displaystyle 4}$Computer Science \& Engineering, Michigan State University, East Lansing, MI}
\symbolfootnote[1]{These authors contributed equally.}\
\ \\[-2mm]
{\bf Keywords:} Information, Representation, Cognition, Evolution

\thispagestyle{empty}
\markboth{}{Evolution of representations}
\ \vspace{-0mm}\\
%
\begin{center} {\bf Abstract} \end{center}
Representations are internal models of the environment that can provide guidance to a behaving agent, even in the absence of sensory information. It is not clear how representations are developed and whether or not they are necessary or even essential for intelligent behavior. We argue here that the ability to represent relevant features of the environment is the expected consequence of an adaptive process, give a formal definition of representation based on information theory, and quantify it with a measure $R$. To measure how $R$ changes over time, we evolve two types of networks---an artificial neural network and a network of hidden Markov gates---to solve a categorization task using a genetic algorithm. We find that the capacity to represent increases during evolutionary adaptation, and that agents form representations of their environment during their lifetime. This ability allows the agents to act on sensorial inputs in the context of their acquired representations and enables complex and context-dependent behavior. We examine which concepts (features of the environment) our networks are representing, how the representations are logically encoded in the networks, and how they form as an agent behaves to solve a task. We conclude that $R$ should be able to quantify the representations within any cognitive system, and should be predictive of an agent's long-term adaptive success.

\section{Introduction}
The notion of representation is as old as cognitive science itself~\citep[see, e.g.,][]{chomsky,newell,fodor,johnson-laird, marr,pinker,pitt_2008}, but its usefulness for Artificial Intelligence (AI) research has been doubted~\citep{Brooks_91}. In his widely cited article ``Intelligence without representation", Brooks argued instead for a subsumption architecture where the autonomous behavior producing components (or layers) of the cognitive system directly interface with the world and with each other rather than with a central symbol processor dealing in explicit representations of the environment. In particular, inspired by the biological path to intelligence, Brooks argued that AI research needs to be rooted in mobile autonomous robotics and a direct interaction between action and perception. Echoing Moravec (\citeyear{Moravec1984}), he asserted that the necessary elements for the development of intelligence are mobility, acute vision, and the ability to behave appropriately in a dynamic environment~\citep{Brooks_91}. This architecture achieved insect-level intelligence and Brooks argued that a path to higher level AI could be forged by incrementally increasing the complexity of subsumption architecture.

However, 20 years after advocating such a radical departure from the classical approach to AI, the subsumption approach seems to have stalled as well. We believe that the reason for the lack of progress does not lie in the attempt to base AI research in mobile autonomous robots, but that instead representations (also sometimes called ``internal models", \citep{Craik1943,Wolpertetal1995,Kawato1999}) are key to complex adaptive behavior. Indeed, while representation-free robotics has made some important strides~\citep{Nolfi2002}, it is limited to problems that are not ``representation hungry''~\citep{Clark1997a}, \textit{i.e.}, problems that do not require past information or additional (external) knowledge about the current {\em context}. In addition, the technical difficulty of developing a subsumption architecture increases with the number of layers or subsystems. This problem of subsumption architecture mirrors the difficulties of classic representational AI approaches to build accurate and appropriate models of the world. 

An alternative approach to engineering cognitive architectures and internal models is evolutionary robotics~\citep{NolfiFloreano2000}. Instead of designing the structure or functions of a control architecture, principles of Darwinian evolution are used to create complex networks that interface perception and action in non-obvious and often surprising ways. Such structures can give rise to complex representations of the environment that are hard to engineer and equally hard to analyze (see, \textit{e.g.}~\citealtt{Floreano1996}). Evolved representations provide context, are flexible, and can be readjusted given new stimuli that contradict the current assumptions. Representations can be updated during the lifetime or over the course of evolution and thus are able to handle even new sensory input~\citep{Bongardetal2006}. We argue that as robots evolve to behave appropriately (and survive) in a dynamic and noisy world, representations of the environment emerge within the cognitive apparatus, and are integrated with the perceived sensory data to create intelligent behavior--using not only the current state of the environment but crucially taking into account historical data (memory) as well. 

To test this hypothesis and make internal representations of evolved systems accessible to analysis, we propose a new information-theoretic {\em measure} of the degree to which an embodied agent represents its environment within its internal states and show how the capacity to represent environmental features emerges over thousands of generations of simulated evolution. The main idea is that representations encode environmental features because of their relevance for the cognitive system in question~\citep{Clark_1994}. Hence, for our purposes, representations can be symbolic or sub-symbolic (e.g., neural states) as long as they have a physical basis, \textit{i.e.}, as long as they are encoded in measurable internal states. However, we distinguish representations from sensorial input because sensor inputs cannot provide the same past or external context as internal states. We thus explicitly define representations as that information about relevant features of the environment which is encoded in the internal states of an organism and which goes beyond the information present in its sensors~\citep{Haugeland1991,Clark1997a}. In particular, this implies that representations can, at time, {\em misrepresent}~\citep{Haugeland1991}--unlike information present in sensors, which always truthfully correlates with the environment. To illuminate the functioning of evolved cognitive systems, we show how it is in principle possible to determine what a representation is {\em about}, and how representations form during the lifetime of an agent. We argue that our measure provides a valuable tool to investigate the organization of evolved cognitive systems especially in cases where internal representations are ``epistemically opaque''.

\section{Methods}
\subsection{Information-theoretic measure of representation}

Information theory has been used previously to quantify how context can modulate decisions based on sensory input~\citep{Phillipsetal1994,PhillipsSinger1997,KayPhillips2011}. Here, we present an information-theoretic construction that explicitly takes the entropy of environmental states into account. To quantify representation, we first define the relationship between the representing system and the represented environment in terms of  information (shared, or mutual, entropy). Information measures the correlation between two random variables, while the entropy $H$ is a measure of the uncertainty we have about a random variable in the absence of information (uncertainty is therefore potential information).
For a random variable $X$ that can take on the states $x_i$ with probabilities $p(x_i)=P(X=x_i)$, the entropy is given by ~\citep{shannon}:
 \begin{equation}
 \label{H}H(X)=-\sum_{i=1}^{N}p(x_i)\log p(x_i)\;, 
 \end{equation}
where $N$ is the number of possible states that $X$ can take on. 

The information between two random variables characterizes how much the degree of order in one of the variables is predictive of the regularity in the other variable. It can be defined using entropy as the difference between the sum of the entropies of two random variables $X$ and $Y$ [written as $H(X)$ and $H(Y)$] and the joint entropy of $X$ and $Y$, written as $H(X,Y)$:
\begin{equation}
I(X:Y) = H(X) + H(Y) - H(X,Y) = \sum_{xy} p(x,y)  \log \frac{p(x,y)}{p(x)p(y)}\;. \label{info1}
\end{equation}
In Eq.~(\ref{info1}), $p(x)$ and $p(y)$ are the probability distributions for the random variables $X$ and $Y$ respectively [that is, $p(x)=P(X=x)$],
while $p(x,y)$ is the joint probability distribution of the (joint) random variable $XY$. The shared entropy $I(X:Y)$ can also be written in terms of a difference between unconditional and conditional entropies, as
\be
I(X:Y)=H(X)-H(X|Y)=H(Y)-H(Y|X)\;. \label{info2}
\ee
This definition reminds us that information is that which reduces our uncertainty about a system. In other words, it is that which allows us to make predictions about a system with an accuracy that is higher than when we did not have that information. In Eq.~(\ref{info2}), we introduced the concept of a conditional entropy~\citep{shannon}. For example, $H(X|Y)$ (read as ``$H$ of $X$ given $Y$") is the entropy of $X$ when the state of the variable $Y$ is known, and is calculated as
\be
H(X|Y)=-\sum_{xy}p(x,y)\log p(x|y)\;,\label{cond}
\ee  
using the conditional probability $p(x|y)=p(x,y)/p(y)$.

In general, information is able to detect arbitrary correlations between signals or sets of events. We assume here that such correlations instantiate semiotic or information relationships between a representing and represented, and use mutual information to measure the correlation between a network's internal states and its environment [see also~\cite{marstaller}]. So, for example, we could imagine that $X$ stands for the states of an environment, whereas $Y$ is a variable that {\em represents} those states of the environment. We need to be careful, however, to exclude from possible representational variables those that are mere images of the environment, such as the trace that the world leaves in an agent's sensors.  Indeed, mere correlations between internal states and the environment are not sufficient to be treated as representational because they could be due to behavior that is entirely reactive~\citep{Clark1997a}. Haugeland, for example, understands representation as something that  ``stands in" for something in the environment, but that is no longer reflected in the perceptual system of the agent~\citep{Haugeland1991}. Indeed, representation should be different from a mere translation: Consider a digital camera's relationship with its environment. The photo chip guarantees a one-to-one mapping between the environment structure and the camera's state patterns. But a camera is not able to adapt to its environment. By taking a picture, the camera has not `learned' anything about its environment that will affect its future state. It simply stores what it received through its inputs without extracting information from it, \textit{i.e.}, the camera's internal states are fully determined by its sensor inputs. Representation goes beyond mere translation because the content, \textit{i.e.} which feature of the environment is represented, depends on the goals of the system. Not everything is represented in the same way. A camera does not have this functional specification of its internal states.

\begin{figure}
\begin{center}
   \includegraphics[width=4in]{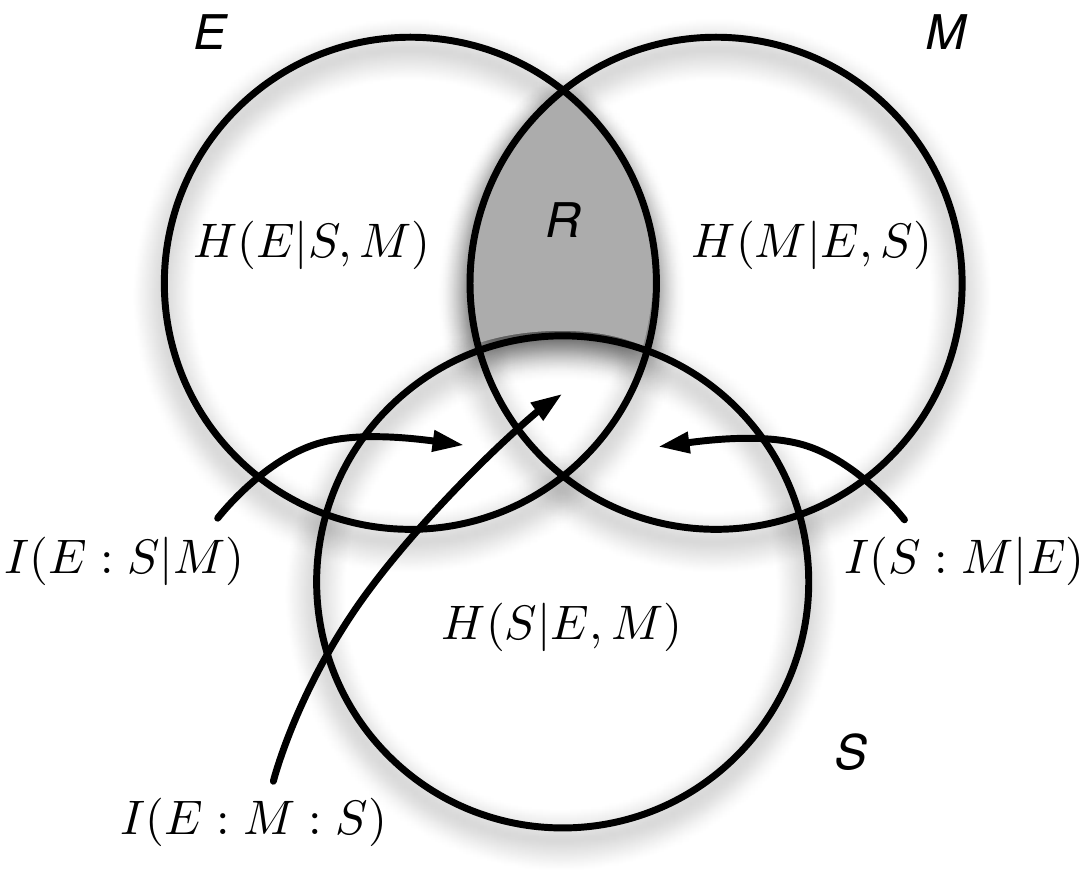} 
\end{center}
\caption{Venn diagram of entropies and informations for the three random variable $E$, $S$, and $M$, describing the states of the environment, sensors, and agent internal degrees of freedom. The representation $R=H(E:M|S)$ is shaded.}
\label{fig:RVenn}
\end{figure}

To rule out trivial representations like a camera's internal states, we define representation as the shared entropy between environment states and internal states, but {\em given} the sensor states, \textit{i.e.}, conditioned on the sensors. Thus, representation is that part of the shared entropy between environment states and internal states that goes beyond what is seen in the sensors (see Fig.~\ref{fig:RVenn}). For the following, we take $E$, given by its probability distribution $p(e_i)=P(E=e_i)$, as the random variable to describe environmental states, while $S$ describes sensor states. If the internal states of the agent (hidden and output states) are characterized by the random variable $M$ with probability distribution $p(m_j)=P(M=m_j)$, then we define the representation $R$ as (for an earlier version, see~\citealtt{marstaller}):
\be
 R=H(E:M|S)=I(E:M)-I(E:M:S)=H_{\rm corr}-I(S:E)-I(S:M)\;, \label{rep}
 \ee
where the correlation entropy $H_{\rm corr}$ of the three variables $E$, $S$, and $M$ [also called ``total correlation"~\citep{Watanabe1960} or ``multi-information"~\citep{McGill1954b,Schneidmanetal2003a}] is the amount of information they all three share:
 \be
 H_{\rm corr} =H(E)+H(M)+H(S)-H(E,M,S)\;.
\ee
In Eq.~(\ref{rep}), we introduced the shared conditional entropy between three variables that is defined as the difference between an information that is unshared and one that is shared (with a third system), just as $H(X|Y)=H(X)-I(X:Y)$, from Eq.~(\ref{cond}).
Thus, the representation $R$ of the world $E$ within internal states $M$ is the total correlation between the three, but without what is reflected in $S$ about $E$ and $M$, respectively [measured by $I(S:E)$ and $I(S:M)$]. The relationship between $R$ and the entropies of the three variables $S$, $E$, and $M$ is most conveniently summarized by an entropy Venn diagram, as in Fig.~\ref{fig:RVenn}. In these diagrams, a circle is a quantitative measure of the entropy of the associated variable, and the shared entropy between two variables is represented by the intersection of the variables, and so on (see,  \textit{e.g.},~\citealtt{cover-thomas-91}).

Our information-theoretic definition of representation carries over from discrete variables to continuous variables unchanged, as can be seen as follows. Let $X_E$, $X_M$, and $X_S$ be random variables defined with normalized probability density functions $f_E(e)$, $f_M(m)$, and $f_S(s)$.
The {\em differential entropy} $h(X)$~\citep{cover-thomas-91}, defined as
\be
h(X)=-\int_{\cal S} f(x)\log f(x)dx\;
\ee
where $\cal S$ is the support of random variable $X$,
is related to the discretized version $H(X^\Delta)$ by noting that
\be
H(X^\Delta)+\log \Delta\to h(X)\;\;\;\;\; (\Delta\to 0)\;,
\ee
where we introduced the discretization
\be
p_i=\int_{i\Delta}^{(i+1)\Delta}f(x)dx
\ee
in order to define the discretized Shannon entropy $H(X^\Delta)=-\sum_ip_i\log p_i$. This implies that an $n$-bit quantization of a continuous random variable is approximately $H(X)\approx h(X)+n$~\citep{cover-thomas-91}. Let us now assume that the variables $X_E$, $X_M$, and $X_S$ are each quantized by $n_E$, $n_M$, and $n_S$ bits respectively. Because $H(E,M,S)$ is then quantized by $n_E+n_M+n_S$ bits, it follows that $H_{\rm corr}\approx h_{\rm corr}$, that is, the continuous and discrete variable correlation entropies are (in the limit of sufficiently small $\Delta$) approximately the same because the discretization correction cancels. The same is true for the informations $I(S:E)$ and $I(S:M)$, as these are correlation entropies between two variables. Thus, $H(E:M|S)\approx h(X_E:X_M|X_S)$, the differential entropy version of $R$. We stress that while an exact identity between discrete and continuous variable definitions of $R$ is only ensured in the limit of vanishing discretization, the cancellation of the correction terms $\log \Delta$ implies that the discrete version is not biased with respect to the continuous version. 

$R$ defines a relation between a network's activity patterns and its environment as the result of information processing. $R$ yields a positive quantity, measured in bits (if logarithms are taken to base 2). In order to show that this measure of representation reflects {\em functional purpose}~\citep{Clark1997a}, we evolve cognitive systems (networks) that control the behavior of an embodied agent, and show that fitness, a measure for the agent's functional prowess, is correlated with $R$. In other words, we show that when the environment (and task) is complex enough, agents react to this challenge by evolving representations of that environment.

\subsection{Evolution of Active Categorical Perception}
\begin{figure}
\begin{center}
  \includegraphics[width=5in]{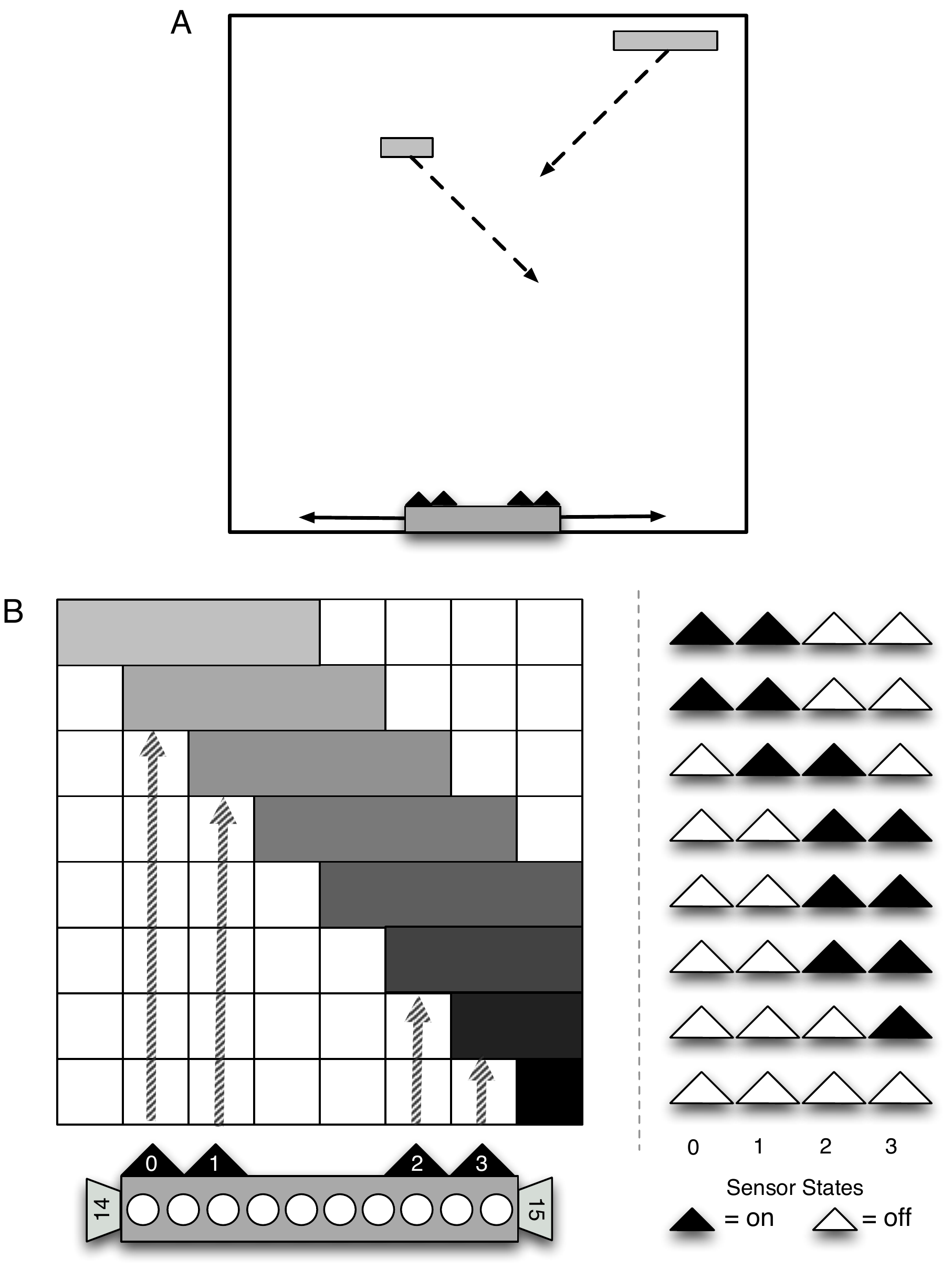} 
\end{center}
\caption{A: In the simulation, large or small blocks fall diagonally towards the bottom row of a $20\times20$ world, with the agent on the bottom row. For the purpose of illustrating the task, a large brick (to be avoided) is falling to the left, while a small brick (to be caught) is falling to the right. In simulations, only one block is falling at the time, and both small and large bricks can fall either to the left or to the right. B:  A depiction of the agent's neurons (bottom left: triangles depict sensors, circles illustrate brain neurons, trapezoids denote actuators) and the sequence of activity patterns on the agent's 4-bit retina (right), as a large brick falls to the right. 
}
\label{fig:task}
\end{figure}

We study the evolution of an agent that solves an active categorical perception (ACP) task~\citep{Beer1996,Beer2003}, but with modifications suggested by \cite{vandarteletal05} (see also~\citealtt{vandarteldiss}). Categorization is thought to be one of the key elements in cognition (see~\citealt{Harnad1987,CohenLefebvre2005}). In categorical perception, an agent has to partition objects in the world into different discrete categories based on their visual appearance. In active categorical perception, the agent takes an active role in which aspects of the object to view, that is, perception is intimately linked with action. Whether or not this task requires internal representations may depend on the specific nature of the task, and it is in general not simple to determine whether the agent uses internal states to represent the environment, and in particular which features of the world are represented by which computational units~\citep{Beer2003,WardWard2009}. In the task studied by Beer~(\citeyear{Beer1996,Beer2003}), the agent has to discriminate between circles and diamonds that were falling vertically towards the agent, who could move laterally to change its perception of the object. In the version we study here, the agent has to discriminate between large and small blocks, by catching the small blocks while avoiding the large ones. In order to create visual ambiguity, the agent's four upward-looking sensors have a blind spot (two non-sensing units) between them.  Furthermore, blocks are falling diagonally (from left to right or right to left), so that the agent not only has to categorize, but also to predict~(see Fig.~\ref{fig:task}A). 

We evolve active categorical perception in two experiments, using two different control architectures. One is an artificial neural network (ANN), and one is a Markov network, specifically, a network of Hidden Markov Gates (HMGs)~\citep{Edlundetal2011}, described in more detail in section~\ref{sec:hmg}.
The agent is located at the bottom row of a $20\times20$ world with periodic boundary conditions, and is able to move horizontally one unit per time step (see Fig.~\ref{fig:task}A). Note that compared to van Dartel et al.~(\citeyear{vandarteletal05}) we doubled the vertical size of the arena in order to give the agent more time to assess the direction of the falling block. The agent has four upward-looking sensors. The state of a sensor is $s_i=+1$ when a block is detected, and $s_i=0$  if no block is seen.
The arrangement of the agent's sensors does not allow it to unambiguously identify the falling block from three out of eight possible input states (counting the input where no sensors are active). Only the input states $S=(s_0,s_1,s_2,s_3)=(0,1,1,0)$, $(0,1,0,0)$, and $(0,0,1,0)$ are unambiguous. The first classifies a large block that is positioned right above the agent so that the second and the third sensor are in state $s_1=s_2=1$ while the remaining two units of the block are positioned above the two non-sensing units. The other two patterns identify similarly positioned small blocks. All other five input states can be created by either a small block, a big block, or no block at all (see Fig.~\ref{fig:task}B). At each trial, a block of either small (two units) or large size (four units) falls from top to bottom in 20 time steps. The blocks move continuously downwards and sideways one unit per time step. Blocks either always move to the right or to the left. An object is caught if the position of the block's units and of the agent's units at time step 20 overlap in at least one unit.

For the information-theoretic characterization of correlations, we have to assign probabilities to the possible states of the world. Theoretically, a falling block can be in any of 20 different starting positions, large or small, and falling left or right, giving rise to 80 possible experimental initial conditions. While the agent can be in any of 20 initial positions, the periodic boundary conditions ensure that each of them is equivalent, given the 20 initial positions of the falling block. Because there are 20 time steps before the block reaches the bottom row, there are in total 1,600 possible different states the world can be in. We do not expect that all of these states will be discriminated by the agent, so instead we introduce a coarse-graining of the world by introducing four bits that we believe capture salient aspects of the world. We define the environmental (joint) variable $E=E_0E_1E_2E_3$ to take on states as defined in Table~\ref{tab:E-states}. 
\begin{table}[htbp]
   \centering
   \begin{tabular}{|@{} |cl| @{}|} 
  \hline
       World  state    & World character\\
      \hline
      $E_0=0$    & no sensor activated \\
    $E_0=1$    & at least one sensor activated \\
   $E_1=0$ & block is to the left of agent\\
   $E_1=1$ & block is to the right of agent \\
   $E_2=0$ & block is two units (small)\\
  $E_2=1$ & block is four units (large)\\
  $E_3=0$ & block is moving left\\
  $E_3=1$ & block is moving right\\
      \hline
   \end{tabular}
   \caption{Coarse-graining of world states into the four bits $E_0,E_1,E_2,E_3$. Note that $E_1$ could be ambiguous in case the block is centered over the agent or exactly 10 units away. We resolve this ambiguity by setting $E_1=0$ when the block is centered over the agent, and $E_1=1$ when it is exactly 10 units away.}
   \label{tab:E-states}
\end{table}
Of course, this encoding reveals a bias in what we, the experimenters, believe are salient states of the world, and certainly underestimates the amount of ``discoverable" entropy. However, in hindsight this coarse-graining appears to be sufficient to capture the essential variations in the world, and furthermore lends itself to study which aspects of the world are being represented within the agent's network controller, by defining representations about different aspects $i$ of the world as the representation $R_i=H(E_i:M|S)$. Thus, we will study the four representations
\be
R_{\rm hit}&=&H(E_0:M|S) \label{rhit}\\
R_{\rm LR}&=&H(E_1:M|S)\label{rlr}\\
R_{4/2}&=&H(E_2:M|S) \label{r42}\\
R_{+/-}&=&H(E_3:M|S)\label{rpm}
\ee
that represent whether the sensor has been activated [Eq.~(\ref{rhit})], whether the block is to the left or the right of the agent [Eq.~(\ref{rlr})], 
if the block is large (size 4) or small (size 2) [Eq.~(\ref{r42})], or whether the block is moving to the left or right [Eq.~(\ref{rpm})]. We can also measure how much (measured in bits) of each binary concept is represented in any particular variable. For example, $R_{\rm LR}({\rm node\ 12})=H(E_1:M_{12}|S)$ measures how much of the ``block is to my left or to my right" concept is encoded in variable 12.

\subsection{Two Architectures for Cognitive Systems}

\label{sec:hmg}
The agent is controlled by a cognitive system, composed of computational units (loosely referred to as ``neurons" from here on) that map sensor inputs into motor outputs. The cognitive system also has neurons that are internal (a hidden layer), which are those neurons that are not part of the input or of the output layer. We further define sensor neurons as those neurons that directly process the input (the input layer) and we define output neurons as those units that do not map to other units in the network or to themselves (the output layer).

\noindent {\bf Artificial Neural Networks (ANN) with evolvable topology.} In our first experiment, the robot's movements are controlled by an artificial neural network that consists of 16 nodes: four input units (one for each sensor), two output units, and ten hidden units. The states of the input units are discrete with values $[+1, -1]$ specifying whether an object is detected or not. The states of the output units (or actuators) are discrete with integer values  $A=a\in [0,1]$ encoding one of three possible actions: move one unit to the right or left, or do not move ($A_1A_2=00: {\rm stand\ still}, A_1A_2=01: {\rm move\ right}, A_1A_2=10:  {\rm move\ left}, A_1A_2=11: {\rm stand\ still}$.
While the hidden units' states $m_i$ are continuous with values $[-1.0,1.0]$, when evaluating these states to calculate $R$ we discretize them to binary (values below $0.0$ become $0$, every other value becomes $1$). As discussed earlier, this discretization does not introduce a bias in the value of $R$. 

Usually, classic artificial neuronal networks have a fixed topology, \textit{i.e.}, one or more layers and their connections are defined and associated with a weight. In a previous experiment, we found that such fixed topologies lead to approximately constant $R$ even as the fitness of the agent increases (data not shown). One way to increase the complexity of the network and the information it represents is to evolve a network's topology as well as the connection weights. We make the network topology evolvable beyond searching the connection weights by using \textit{neuronal gates} (NG). A NG can arbitrarily connect nodes of any type (input, hidden, and output nodes) without the fixed layered topology of classic ANNs. Each connection is associated with a certain weight. A NG calculates the sum of the values from a set of incoming nodes $m^{(k)}_j$ via gate $k$, multiplied by the associated weight $w^{(k)}_{j}$ and applies a sigmoid function to calculate its output $m$ 

\begin{equation}
m(t+1) = \tanh\left( \sum_{k=1}^n \sum_{j=1}  w^{(k)}_{j} m^{(k)}_j(t)\right)\;,\label{theta}
\end{equation}
where the sum over $j$ runs over all the neurons that feed into gate $k$. This value $m$ is then propagated to every node this NG is connected to. 

\begin{figure}
\begin{center}
  \includegraphics[width=4.5in]{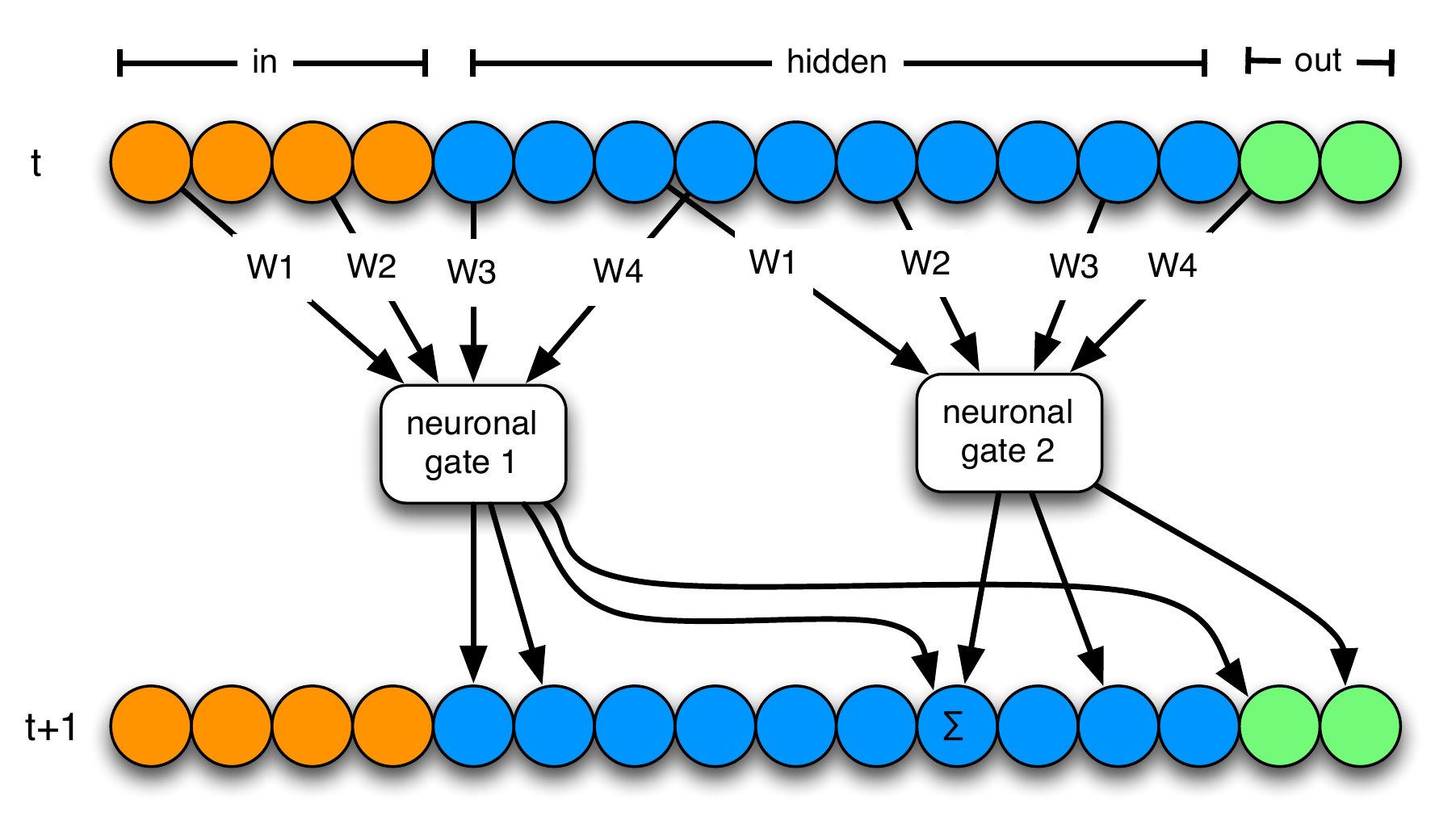}  
\end{center}
\caption{Open-topology Artificial Neural Network, with four input neurons (orange), ten hidden neurons (blue), and two motor neurons (green). The nodes are connected via two neuronal gates (NG). Each NG connects four arbitrary input nodes with weight $W$ to four output nodes. This figure illustrates how the nodes become updated from time point $t$ to $t+1$. When two NGs write into the same node, their outputs are added (indicated by $\sum$) before the sigmoid function is applied.}
\label{fig:network}
\end{figure}

To apply a Genetic Algorithm to this system, each ANN is encoded in a genome as follows: A start codon of two loci mark the beginning of a NG, the subsequent two loci encode the NG's number of inputs and outputs, while two further loci specify the origin of the inputs (which neurons feed into the gate) and the outputs (where the NG writes into). This information is then followed by an encoding of the $n$ weights of an $n$-input NG (see Fig.~\ref{fig:network}). The number of gates in the network can change as it evolves, and is only determined by the number of start codons in the genome. The genomes encoding these ANNs can undergo the same mutational changes as described later in the MB section. In this respect, evolving open-topology ANNs is similar to using evolutionary algorithms (such as NEAT, see, \textit{e.g.},~\cite{Stanley2002}) to evolve neural networks with augmented topology. 

\noindent {\bf Markov Brains (MB).} In our second experiment, the agent is controlled by a network of 16 nodes (four input, two output, and ten internal nodes, \textit{i.e.}, with the same types and number of nodes as the ANNs) which are connected via \textit{Hidden Markov Gates} (HMGs, see~\citealt{Edlundetal2011}). Networks of HMGs (Markov brains or MBs for short) are a type of stochastic Markov network (see,  \textit{e.g.},~\citealt{KollerFriedman2009}) and are related to the hierarchical temporal memory model of neocortical function~\citep{HawkinsBlakeslee2004,GeorgeHawkins2005,GeorgeHawkins2009} and the HMAX algorithm~\citep{RiesenhuberPoggio1999}, except that Markov brains need not be organized in a strictly hierarchical manner as their connectivity is evolved rather than designed top-down.

\begin{figure}[htp]
\begin{centering}
\includegraphics*[width=14cm, clip=true]{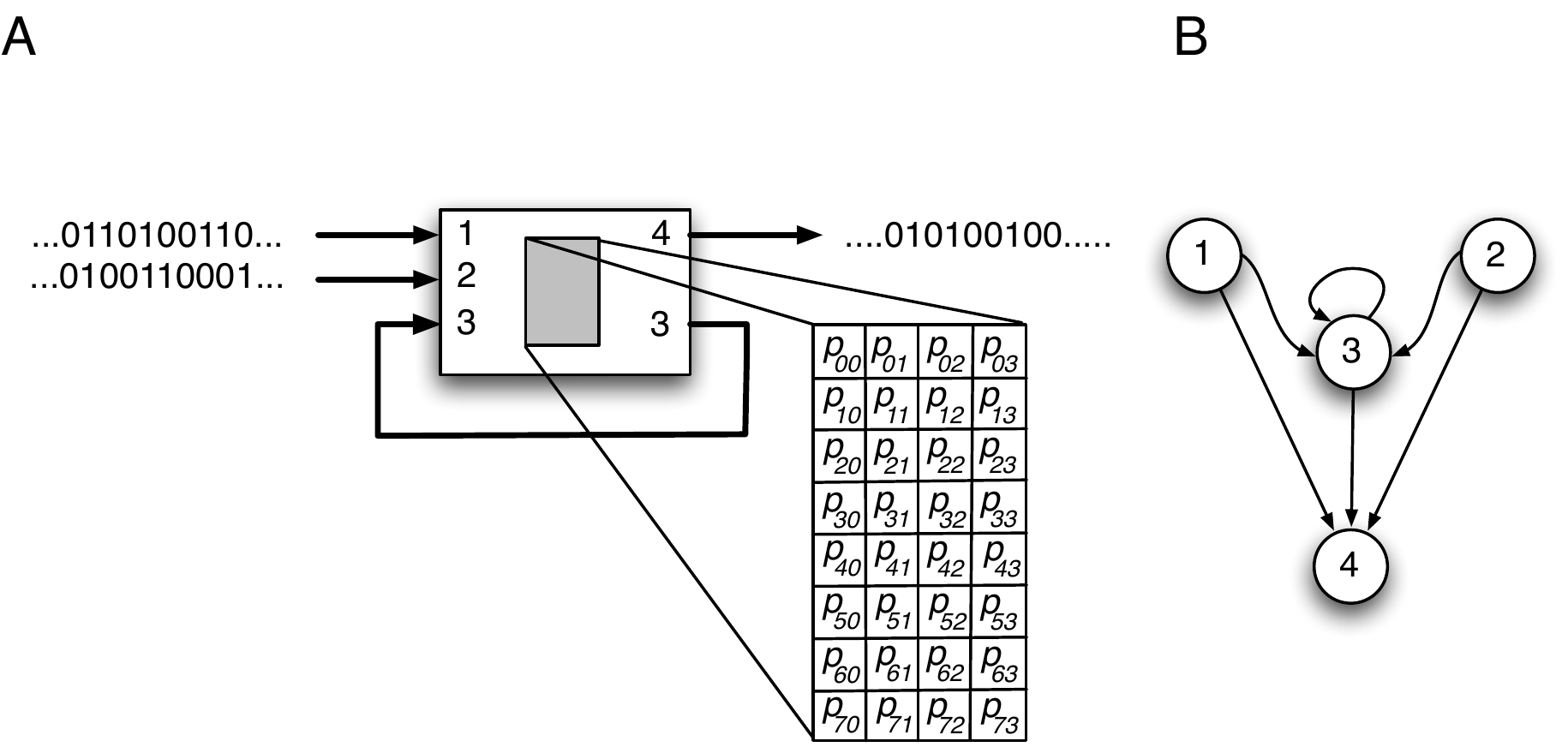}
\caption{{\bf A}:  A single HMG with three inputs and two outputs reads from nodes 1, 2, and 3, and writes to nodes 3 and 4, updating the states of these nodes in the process. {\bf B}: The output states are determined by a set of $2^{3+2}$ probabilities, here denoted as $p_{xy}$, where $x$ and $y$ are the decimal equivalent of the binary pattern of the input and output, respectively. For example, $p_{73}$ is the probability for the pattern `11' to fire if the input was `111', that is, $P(11|111)=p_{73}$.}
\label{fig:gates}
\end{centering}
\end{figure}

Each HMG can be understood as a finite-state machine that is defined by its input/output structure (Fig. \ref{fig:gates}A) and a state transition table (Fig.~\ref{fig:gates}B). All nodes in Markov brains are binary, and in principle the HMGs are stochastic, that is, the output nodes fire (that is, are set to state `1') with a probability determined by the state-to-state transition table. Here, each HMG can receive up to four inputs, and distribute signals to up to 4 nodes, with a minimum of one input and one output node (these settings are configurable). For the evolution of the ACP task, we consider only {\em deterministic} HMGs (each row of the transition table contains only one value of 1.0 and all other transitions have a probability of 0.0), turning our hidden Markov gates into classical logic gates. In order to apply an evolutionary algorithm, each HMG is encoded in a similar way as the NGs using a genome that specifies the network as a whole. Each locus of the genome is an integer variable $\in[0,255]$. Following a start codon (marking the beginning of a gene, where each gene encodes a single HMG), the next two loci encode the number of inputs and outputs of the gate respectively, followed by a specification of the origin of the inputs, and the identity of the nodes being written to. 

For example, for the HMG depicted in Fig.~\ref{fig:gates}, the loci following the start codon would specify `3 inputs', `2 outputs', 'read from 1,2,3', `write to 3,4'. This information is then followed by an encoding of the $2^{n+m}$ probabilities of an $n$-input and $m$-output state transition table (see Supplementary Fig. S1 in~\citealt{Edlundetal2011} for more details.) For the example given in Fig.~\ref{fig:gates}, the particular HMG is specified by a circular genome with 39 loci (not counting the start). The start codon is universally (but arbitrarily) chosen as the consecutive loci (42,213). Because this combination only occurs by chance once every 65,536 pairs of loci (making start codons rare), we insert four start codons at arbitrary positions into a 5,000 loci initial genome to jump start evolution. Thus, the ancestral genomes of all experiments with Markov brains encode at least 4 HMGs. A set of HMGs encoded in this manner uniquely specifies the Markov brain. The encoding is robust in the sense that mutations that change the input-output structure of an HMG leave the probability table intact, while either adding or removing parts of the table. 
This flexibility also implies that there is considerable neutrality in the genome, as each gene has 256 loci reserved for the probability table even if many fewer loci are used.

MBs and ANNs differ with respect to the gates connecting the nodes in each network. ANNs use weights, sums, and a $\tanh$ function together with continuous variables to compute their actions. In contrast MBs use discrete states and boolean logic to perform their computation. Using a very similar encoding of the topology means that mutations will have a similar effect on the topology of both systems, but different effects on the computations each system performs. 

\subsection{Evolutionary Algorithm}
\label{sec:ea}
We evolve the two types of networks (ANNs and MBs) using a Genetic Algorithm (GA). A GA can find solutions to problems by using evolutionary search [see, e.g., \cite{Michalewicz1996}]. The GA operates on the specific genetic encoding of the networks' structure (the genotype), by iterating through a cycle of assessing each network's fitness in a population of 100 candidates, selecting the successful ones for differential replication, and finally mutating the new candidate pool. When testing a network's performance in controlling the agent, each network is faced with all 80 possible initial conditions that the world can take on. The fitness $w$ is calculated as the fraction of successful actions (the number of large blocks avoided plus the number of small blocks caught) out of 80 tests (a number between zero and 1). For the purpose of selection, we use an exponential fitness measure that multiplies the score by a factor 1.1 for every successful action, but divides the score by 1.1 for every unsuccessful action, or $S= 1.1^{80(2w-1)}$. After the fitness assessment,  the genotypes are ranked according to $S$ and placed into the next generation with a probability that is proportional to the fitness (roulette wheel selection without elite). After replication, genotypes are mutated.
We implemented three different mutational mechanisms that all occur after replication, with different probabiities. A point mutation happens with a probability of $\mu=0.005$ per locus, and causes the value at that locus to be replaced by a uniform random number drawn from the interval $[0\cdots 255]$. There is a 2\% chance that we delete a sequence of adjacent loci ranging from 256-512 in size, and a 5\% chance that a stretch of 256-512 adjacent loci is duplicated (the size of the sequence to be deleted or duplicated is unformly distributed in the range given). The duplicated stretch is randomly inserted between any two loci in the genome.  Duplications and deletions are contrained so that the genome is not allowed to shrink below 1024 sites, and genomes cannot grow beyond 20,000 sites. Because insertions are more likely than deletions, there is a tendency for genomes to grow in size during evolution. 

We evolve networks through 10,000 generations, and run 200 replicates of each experiment. Note that the type of gates is different between ANNs (neuronal) and MBs (logic), so the rate of evolution of the two networks cannot be compared directly, because mutations will have vastly different effects with respect to the function of the gates. Thus, the optimal mutation rate differs among networks~\citep{Orr2000}.
At the end of each evolutionary run, we reconstruct the evolutionary {\em line of descent}~\citep{Lenskietal2003} of the experiment, by following the lineage of the most successful agent at the end of 10,000 generations backwards all the way to the random ancestor that was used to seed the experiment. This is possible because we do not use cross-over between genotypes in our GA. This line of descent, given by a temporally ordered sequence of genotypes,  recapitulates the unfolding of the evolutionary process, mutation by mutation, from ancestor to the evolved agent with high fitness, and captures the essence of that particular evolutionary history. For each of the organisms on each of the 200 lines of descent of any particular experiment, we calculate a number of information-theoretic quantities, among which is how much of the world the agent represents in its brain, using Equation~(\ref{rep}). 

\subsection{Extracting probabilities from behavior} \label{sec:prob}
The inputs to the information-theoretic measure of representation $R$ are the probabilities to observe a particular state $x$, $p(x)$, as well as the joint probabilities $p(x,y)$ describing the probability to observe a state $x$ when at the same time another variable $Y$ takes on the state $Y=y$. For the representation $R$ defined by Eq.~(\ref{rep}), sensor, internal, and environment variables are distinguished. For any particular organism (an agent that performs the ACP task with an evolved controller), $R$ is measured at any point during the evolution by placing the organism into the simulated world and concurrently recording time series data of the states of all 16 controller nodes and the states of the environment. The recordings are then used to calculate the frequency of states. Based on the frequencies of states, all probabilities relevant for the information-theoretical quantities can be calculated, including those that take into account the temporal order of events (for example, the probability $p(x_t,y_{t+1})$ that variable $X_t$ takes on state $x_t$ while variable $Y_{t+1}$ takes on the state $y_{t+1}$). If a particular state (or combination of states) never occurs, a probability of zero is recorded for that entry (even though in principle the state or combination of states could occur). $R$ (and the other information-theoretic quantities introduced in section 3) is calculated for organisms on the evolutionary line of descent making it possible to follow the evolutionary trajectory of $R$ from random ancestor to adapted agent. For ANNs that have internal nodes with continuous rather than binary states, a mapping of intervals $[-1,0]\to0$ and $]0,1]\to1$ is applied before calculation of probabilities.

\section{Results}
To establish a baseline, 100,000 random controllers for each of the two network types were created and the distributions of $R$ and fitness values were obtained. This baseline served two purposes: it shows how well randomly generated (unevolved) networks perform, and how much information about the world they represent by chance as well as providing information about the distribution of these values. Random ANNs and MBs were created in the same way by randomly drawing values from a uniform probability distribution of the integers $\in[0,255]$ for each of the genome's loci. Each genome was then sprinkled with 4 start codons at arbitrary positions within the genome.

\begin{figure}[htbp]
\centering
\includegraphics[width=5.5in]{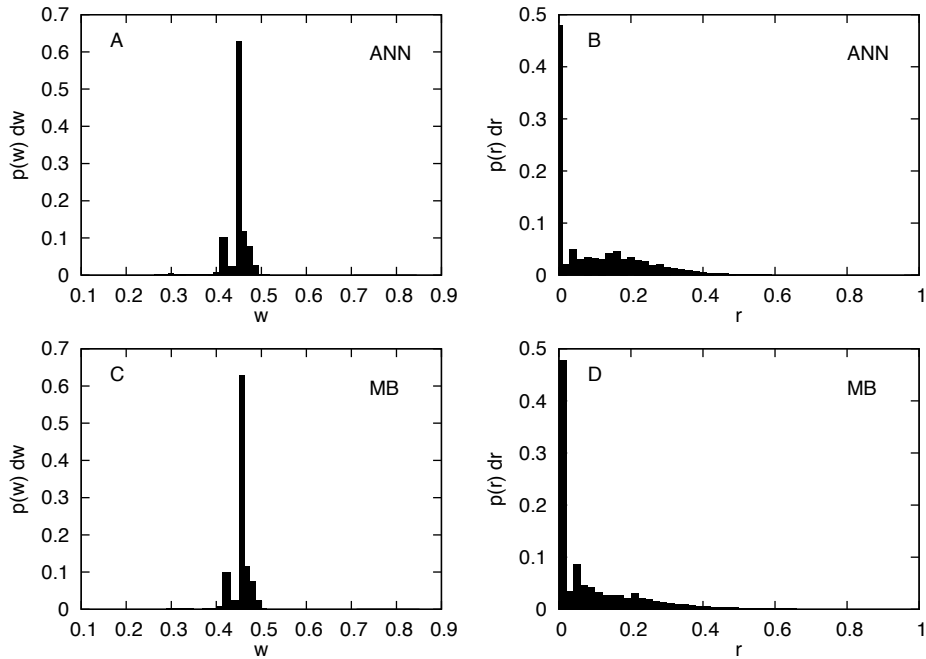}
\caption{Probability distribution of fitnesses and representation scores of random machines. {\bf A}:  Probability distribution of fitnesses (fraction of successful actions) $p(w)dw=P(w<W<w+dw)$ for 100,000 random ANNs ($dw=1/80$) {\bf B}: Probability distribution of the representation variable $R$ for the same random ANNs $p(r)dr = P(r<R<r+dr)$, with $dr=0.02$. {\bf C}: Distribution of fitnesses for 100,000 random Markov brains ($dw=1/80$). {\bf D}: Distribution of representation $R$ in the same MB networks ($dr=0.02$).}
\label{fig:hist}
\end{figure}

Fig.~\ref{fig:hist}A shows the distribution of fitness scores $w$ for 100,000 random ANNs, and Fig.~\ref{fig:hist}B shows the distribution of their representation $R$. The respecitve distributions for MBs are shown in Fig.~\ref{fig:hist}C for fitness scores and in Fig.~\ref{fig:hist}D for representation. While both systems use different types of gates, ANNs and MBs do not differ with respect to their initial fitness or representation distributions. This shows that random genomes with a high fitness are--as expected--very rare, and need to evolve their functionality in order to perform optimally.

\begin{figure}[htbp]
\centering
\includegraphics[width=\textwidth]{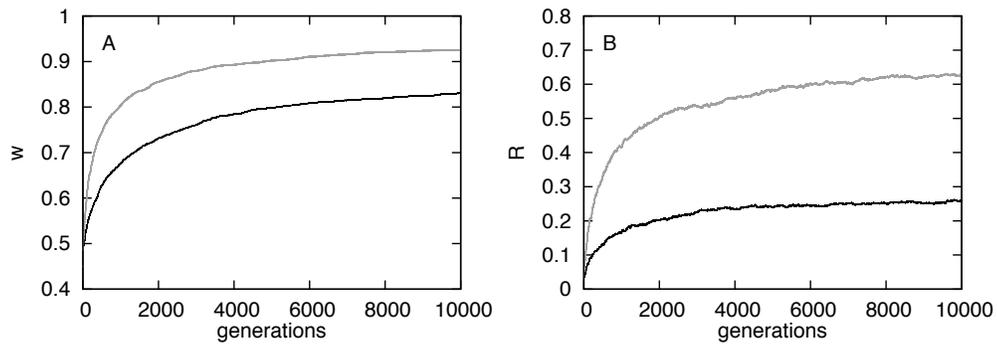}
\caption{A: Fitness $w$ and B: representation $R$ (in bits) along the line of descent as a function of evolutionary generations, averaged over 200 independent evolutionary lines, for evolved networks (ANNs: black, MBs: blue)}
\label{fig:fit-rep}
\end{figure}

In order to compare the two network architectures, the evolutionary trajectories for fitness and representation were analyzed for the evolutionary line of descent (LOD) as described in section \ref{sec:ea}. The different LODs obtained from following back any other member of the final population quickly coalesce to a single line. Hence, the LOD effectively recapitulates the genetic changes that led from random networks to proficient ones. The development of fitness and representation over evolutionary time in Fig.~\ref{fig:fit-rep} is averaged over 200 independent replicates. While both ANNs and MBs have on average low fitness at the begin of an evolutionary run (as seen in Fig.~\ref{fig:hist}), MBs become significantly more fit than ANNs, and after 10,000 generations we find that in 18 out of 200 runs MBs have evolved to perfect fitness, while none of the ANNs reached this level (the best ANNs correctly make $77$ out of $80$ decisions). At the same time, we see that the fitness of the ANNs after 10,000 generations is not stagnating, which suggests that more runtime will allow for further improvement. As previously mentioned, the rate at which fitness is achieved in evolutionary time cannot be compared across architectures because mutations affect the function of the gates differently. While we tentatively explain the difference in performance between ANNs and MBs by their difference in representing the world below, we anticipate that the different network architectures also solve the categorization task very differently. To understand the information dynamics and the strategies employed in more detail, we measured a number of other information-theoretic measures (besides $R$) (see sections below).

The evolutionary trajectory for representation $R$ (see Fig.~\ref{fig:fit-rep}B) is similar to the evolution of fitness (see Fig.~\ref{fig:fit-rep}A), but MBs evolve to a significantly higher value of $R$. We attribute this difference to the difference in fitness between the two types of networks, as the discretization of the continuous ANN variables cannot introduce a bias in $R$. Thus, it appears that an increased representation of the world within an agent's network controller correlates with fitness. We can test this correlation between fitness and representation at the end of a run for the 200 replicates of MBs and ANNs, and find that fitness and $R$ are significantly correlated (Spearman's $r=0.55$, $P=2.5\times10^{-17}$) for MBs, but not for ANNs ($r=-0.10$, $P=0.15$). We speculate that because ANNs are forced to compute using a sigmoid function only (effectively implementing a multiple-AND gate) while MBs can use arbitrary logic operations to process data, ANNs struggle to internalize (that is, represent) environmental states. In other words, it appears that the ease of memory-formation is crucial in forming representations, which are then efficiently transformed into fit decisions in MBs~\citep{Edlundetal2011}.

\subsection{Analysis of Network Structures and Strategies}
In order to be successful at the task described, an agent has to perform active categorical perception followed by prediction. In the implementation of the ACP task by Beer (\citeyear{Beer1996,Beer2003}), prediction can be achieved without memory, because once the network has entered the attractor representing a category, the prediction (to move away or to stay) can be directly coupled to the attractor. The task used here can only be achieved using memory (data not shown) and requires the agent to perform categorical perception by comparing sensory inputs from at least two different time points (which also allows a prediction of where the object is going to land). 

\begin{figure}[htp]
\begin{centering}
\includegraphics*[width=10cm]{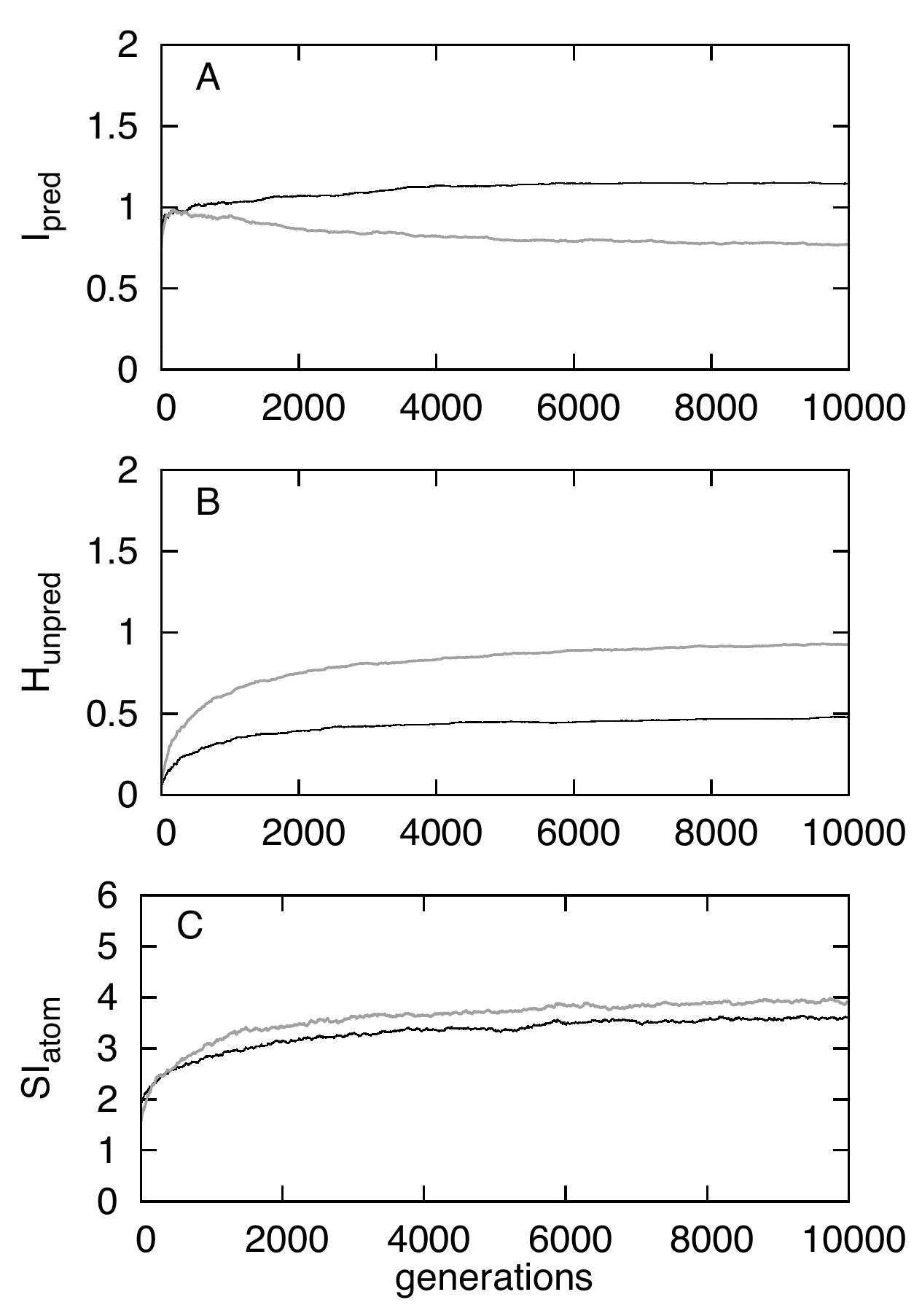}
\caption{Different measures of information processing and integration along the LOD for both types of network architectures: ANNs (black) and Markov brains (blue). A: Predictive information, Eq.~(\ref{Ipred}). B: Unpredicted entropy [Eq.~(\ref{hbrain})] of the network's motor variables. C: Information integration $SI_{\rm atom}$ based on Eq.~(\ref{eqatom}). }
\label{fig:var_meas}
\end{centering}
\end{figure}

In order to analyze how information is processed, we calculated the \textit{predictive information}~\citep{bialek} of the evolved networks, given by the mutual Shannon information between the network's inputs at time $t$ and its outputs at time $t+1$. Predictive information, defined this way~\citep{Ayetal2008}, measures how much of the entropy of outputs (the firings of motor neurons that control the agent) can be understood in terms of the signals that have appeared in the agent's sensors just prior to the action. Indirectly, predictive information therefore also indicates the contribution of the hidden nodes of the network. A high predictive information would show that the hidden nodes do not contribute much and that computations are performed mainly by input and output neurons. Using the variable $S$ for sensor states and $A$ for actuator states, the predictive information can be written in terms of the shared entropy between sensor states at time $t$ and motor states at time $t+1$ as

\begin{equation}
I_{\rm pred}=I(S_t:A_{t+1})=-\sum_{s_t,a_{t+1}}{p(s_t,a_{t+1})\log{\frac{p(s_t,a_{t+1})}{p(s_t)p(a_{t+1})}}} \;,\label{Ipred}
\end{equation}
where $p(s_t)=P(S_t=s_t)$ is the probability to observe variable $S_t$ in state $s_t$, $p(a_{t+1})=P(A_{t+1}=a_{t+1})$ is the probability observe variable $A_{t+1}$ in state $a_{t+1}$, etc.  Note that $S_t$ and $A_{t+1}$ are joint random variables created from the variables of each node, implying that $S_t$ can take on 16 different states while $A_{t+1}$ can take on 4 possible states. The probabilities are extracted from time series data as described in section~\ref{sec:prob}. Figure  \ref{fig:var_meas}A shows that over the course of evolution, the predictive information $I_{\rm pred}$ decreases for MBs after an initial increase, but increases slightly overall for ANNs. The drop in predictive information for the MBs indicates that their actions become less dependent on sensor inputs and are driven more by the hidden neurons, while the ANNs actions remain to be predictable by sensor inputs.

To test whether it is the internal states that increasingly guide the agent, the predictive information was subtracted from the entropy of the output states (maximally two bits) to calculate the \textit{unpredicted entropy} of the outputs, \textit{i.e.}, how much of the motor outputs are uncorrelated to signals from the input:

\begin{equation}
H_{\rm unpred}=H(A_{t+1})-I_{\rm pred}= H(A_{t+1}|S_t)\;.\label{hbrain}
\end{equation}

Figure \ref{fig:var_meas}B shows that $H_{\rm unpred}$ increases over the course of evolution, suggesting that indeed signals other than the sensor readings are guiding the motors. In principle, this increase could be due to an increase in the motor neuron entropy, however, as the latter stays fairly constant we can conclude that the more a network adapts to its environment, the less its outputs are determined by its inputs and the more by its internal states. Again, this effect is stronger for MBs than for ANNs, and suggests that it is indeed the internal states that encode representations that drive the network's behavior. 
It is also possible that the motors evolve to react to sensor signals further back in time. Because sensor neurons cannot store information, such a delayed response also has to be processed via internal states. While the absolute value of the predicted information and unpredicted entropy can depend on this time delay, we expect the overall trend of a decreasing $I_{\rm pred}$ coupled with an increasing $H_{\rm unpred}$ to be the same as for the one-step predictive information, because the sensorial signal stream itself has temporal correlations, that is, it is non-random. 

To quantify the synergy of the network we calculated a measure of information integration called \textit{synergistic information}. Roughly speaking, synergistic information measures the amount of information that is processed by the network as a whole that cannot be understood in terms of the information-processing of each individual node, \textit{i.e.}, it measures the extent to which the whole network is--informationally--more than the sum of its parts~\citep{Edlundetal2011}:

\begin{equation}
SI_{\rm atom}= I(X_t:X_{t+1})-\sum_{i=1}^n{I(X_t^i:X_{t+1}^i)} \;. \label{eqatom}
\end{equation}
In Eq.~(\ref{eqatom}), $I(X_t:X_{t+1})$ measures the amount of information that is processed (across time) by the whole network $X$ (the joint random variable composed of each of the node variables), whereas $I(X_t^i:X_{t+1}^i)$ measures how much is processed by node $i$. The negative of Eq.~(\ref{eqatom}) has been used before, to quantify the {\em redundancy} of information processing in a neural network~\citep{Atick1992,NadalParga1994,Schneidmanetal2003a}. $SI_{\rm atom}$ is a special case of the information integration measure $\Phi$~\citep{tononi_2008, balduzzi}, which is computationally far more complex than $SI_{\rm atom}$ because it relies on computing information integration across all possible partitions of a network. $SI_{\rm atom}$, instead, calculates information integration across the ``atomic" partition only, that is, the partition where each node is its own part.
Figure \ref{fig:var_meas}C shows that $SI_{\rm atom}$ increases for Markov brains as well as for ANNs, which indicates that both architectures evolve the ability to integrate information to perform the task at hand. We only see a marginal difference between MBs and ANNs in their ability to integrate information, while at the same time MBs are more dependent on internal states and ultimately perform better. This suggests that measuring integrated information in terms of Eq.~\ref{eqatom} does not allow inferences about a system's capacity do memorize. In summary, we observe that MBs evolve to become less dependent on sensorial inputs than ANNs, and in addition, the actions of the MBs become more dependent on internal states than in ANNs. Thus, we conclude that the network properties that $R$ measures are indeed representations that the networks create as an adaptive strategy. But how do the networks represent their environments? Which features of the world are represented and form the successful task-solving strategies?

\subsubsection{Epistemically Opaque Strategies}
In order to analyze a Markov brain's function, a number of different tools are available. First, a {\em causal diagram} can be generated by drawing an edge between any two neurons that are connected via an HMG. The edges are directed, but note that each edge can in principal perform a different computation.  When creating the causal diagram, nodes that are never written into by any other nodes are removed, as they are computationally inert (they remain in their default `off' state). Such nodes can also be identified via a ``knock-out" procedure, where the input of each node is forced to either the ``0" or ``1" state individually. If such a procedure has no effect on fitness, the node is inert.

\begin{figure}[htbp]
\begin{centering}
\includegraphics*[width=4in, clip=true]{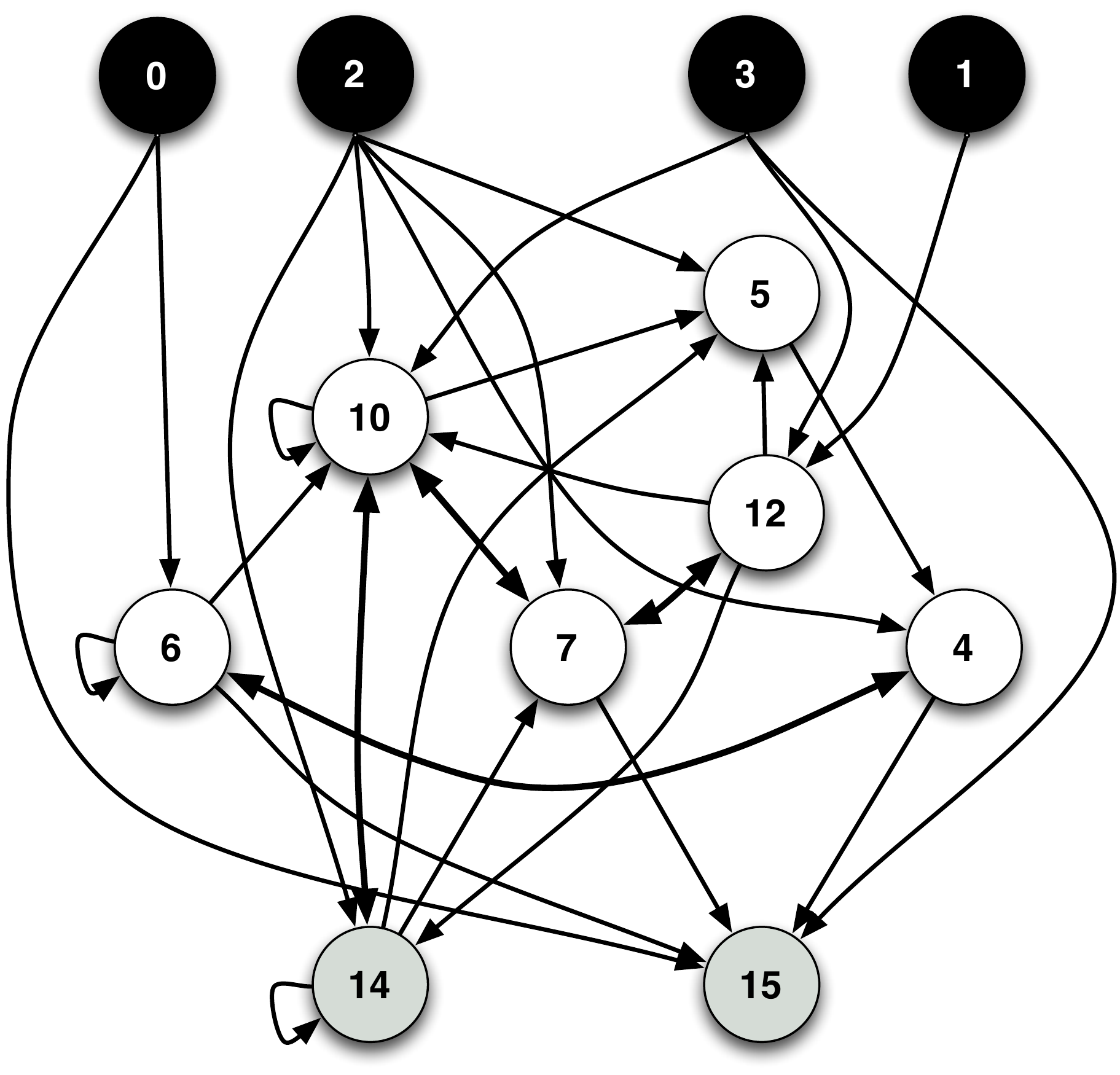}
\caption{Causal diagram of a Markov brain with perfect fitness, correctly catching all small blocks and avoiding all large ones. Nodes colored in red are sensors, while motor variables are green. Double arrows represent two causal connections, one each way. Nodes with arrows that point to themselves write their output back into their input, and may work as memory (all nodes return to a default ``0" if not set otherwise by each update), so state information can only be maintained via such self-connections). Internal nodes can read from motors, giving rise to proprioception, or more precisely, {\em kinesthesia}: the ability to sense one's own motion. The motors themselves can be used as memory. 
\label{fig:perfect}}
\end{centering}
\end{figure}

Fig.~\ref{fig:perfect} shows the causal diagram of an evolved Markov brain that solves the classification task perfectly. One can see that this network uses inputs from the sensors, motors, and memory simultaneously for decisions, fusing the different modalities intelligently~\citep{Murphy1996}.

The causal diagram by itself, however, does not reveal {\em how} function is achieved in this network. As each HMG in the present instantiation represents a deterministic logic gate (generally they are stochastic), it is possible to determine the logical rules by which the network transitions from state to state by feeding the state-to-state transition table into  a logical analyzer ~\citep{logicfriday}. The analyzer converts the state transition table into the minimal description of functions in Boolean logic using only NOT, AND ($\wedge$), and OR ($\vee$). With these functions we can exactly describe each node's logical influence on other nodes (and possibly itself). For the network depicted in Fig.~\ref{fig:perfect}, the logic is given by (here, the numeral represents the node, and its index the state at time $t0$ or the subsequent time point $t1$, while an overbar stands for NOT)
\be
4_{t1} &=& (\bar2_{t0}  \wedge \bar 5_{t0})   
\vee
( 2_{t0} \wedge 6_{t0})\nonumber\\
5_{t1}&=&
(\bar2_{t0} \wedge 10_{t0} \wedge \bar{12}_{t0})
\vee
(\bar{10}_{t0} \wedge \bar{12}_{t0} \wedge \bar{14}_{t0})
\vee
(\bar 2_{t0} \wedge 10_{t0} \wedge 12_{t0}\wedge \bar{14}_{t0})
\nonumber\\
6_{t1} &=&
\bar 4_{t0} \wedge(  0_{t0}  \vee  6_{t0}) 
\nonumber\\
7_{t1} &=& 
(2_{t0} \wedge \bar{12}_{t0})
\vee
(\bar {10}_{t0} \wedge 12_{t0} \wedge 14_{t0})
\nonumber\\
10_{t1}& =& (2_{t0}\wedge6_{t0}) 
\vee
( \bar 2_{t0} \wedge 10_{t0} \wedge \bar {14}_{t0})
\vee
(10_{t0} \wedge 12_{t0} \wedge \bar {14}_{t0}) 
\vee
(2_{t0} \wedge 14_{t0}) \nonumber\\
&&
\vee
(3_{t0} \wedge \bar7_{t0} \wedge \bar {10}_{t0}\wedge12_{t0}\wedge14_{t0} )
\nonumber\\
12_{t1} &=& 
(\bar1_{t0}  \wedge \bar 7_{t0})   
\vee
(\bar3_{t0}  \wedge  7_{t0})   
\vee
(3_{t0}  \wedge \bar 7_{t0})   \nonumber\\
14_{t1} &=& 
(2_{t0} \wedge 10_{t0} \wedge {14}_{t0})
\vee
(2_{t0} \wedge \bar{10}_{t0} \wedge \bar{12}_{t0})
\vee
(\bar 2_{t0} \wedge 10_{t0} \wedge {12}_{t0}) \nonumber \\
15_{t1} &=& 
(3_{t0} \wedge \bar{7}_{t0})
\vee
(\bar 0_{t0} \wedge {6}_{t0})
\vee
( 4_{t0} \wedge {6}_{t0})
\vee
(0_{t0} \wedge \bar 4_{t0} \wedge \bar6_{t0}) \nonumber\;.
\ee
Note that while this logical representation of the network's dynamics is optimized (and the contribution of inert nodes is removed), it is in general not possible to determine the {\em minimal} logic network based on state-to-state transition information only, as finding the minimal logic is believed to be a computationally intractable problem~\citep{KabanetsCai2000}. As a consequence, while it is possible to capture the network's function in terms of a set of logical rules, we should not be surprised that evolution delivers epistemically opaque designs~\citep{Humphreys2009}, that is, designs that we do not understand on a fundamental level. However, the strength of measuring representations with $R$ is that our measure is able to capture representations even if they are highly distributed and take part in complex computations. As such, $R$ provides a valuable tool to analyze evolved neural networks as can be seen below.

\begin{figure}[htbp]
\begin{centering}
\includegraphics*[width=\linewidth, clip=true]{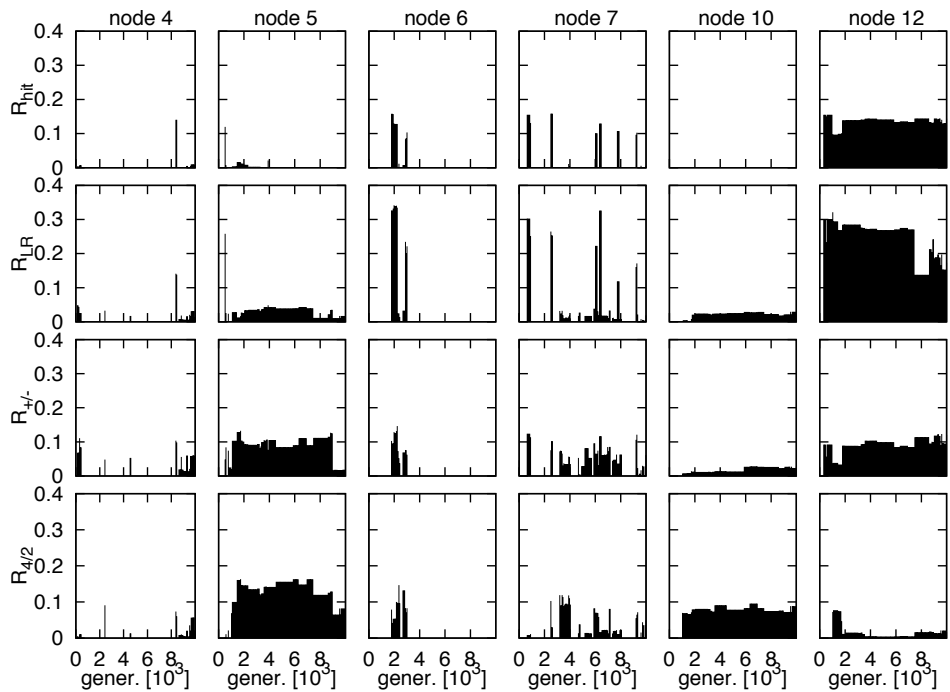}
\caption{Representation of each of the 4 environmental properties ({\em concepts}) defined in Eqs.~(\ref{rhit}-\ref{rpm}) as a function of time, within each of the nodes of a network that evolved to become the one depicted in Fig.~\ref{fig:perfect}. Representation is measured in bits, along the temporal (genetic) line of descent (measured in generations). }
\label{fig:R_node}
\end{centering}
\end{figure}

\subsubsection{Concepts and Memory}
To understand what representations are acquired (representations about which concepts), we calculated $R$ for each property of the environment defined in Eq.~(\ref{rhit}-\ref{rpm}), within each of the key nodes in our example network shown in Fig.~\ref{fig:perfect}. For this network, Fig.~\ref{fig:R_node}
shows that some nodes prefer to represent given single features or concepts,  while others represent several features at the same time. In addition, the degree to which a node represents a certain property changes during the course of evolution. Looking at representation within each individual node, however, only tells part of the story as it is clear that representations are generally ``smeared" over several nodes. If this is the case, a pair of nodes (for example) can represent more about a feature than the sum of the representations in each node, \textit{i.e.}, variables can represent synergistically. In order to discover which combination of nodes represents which feature most accurately, a search over all partitions of the network would have to be performed, much like in the search for the partition with minimum information processing in the calculation of a network's synergistic information processing~\citep{balduzzi}. 

\begin{figure}[htbp]
\centering
\includegraphics[width=\linewidth]{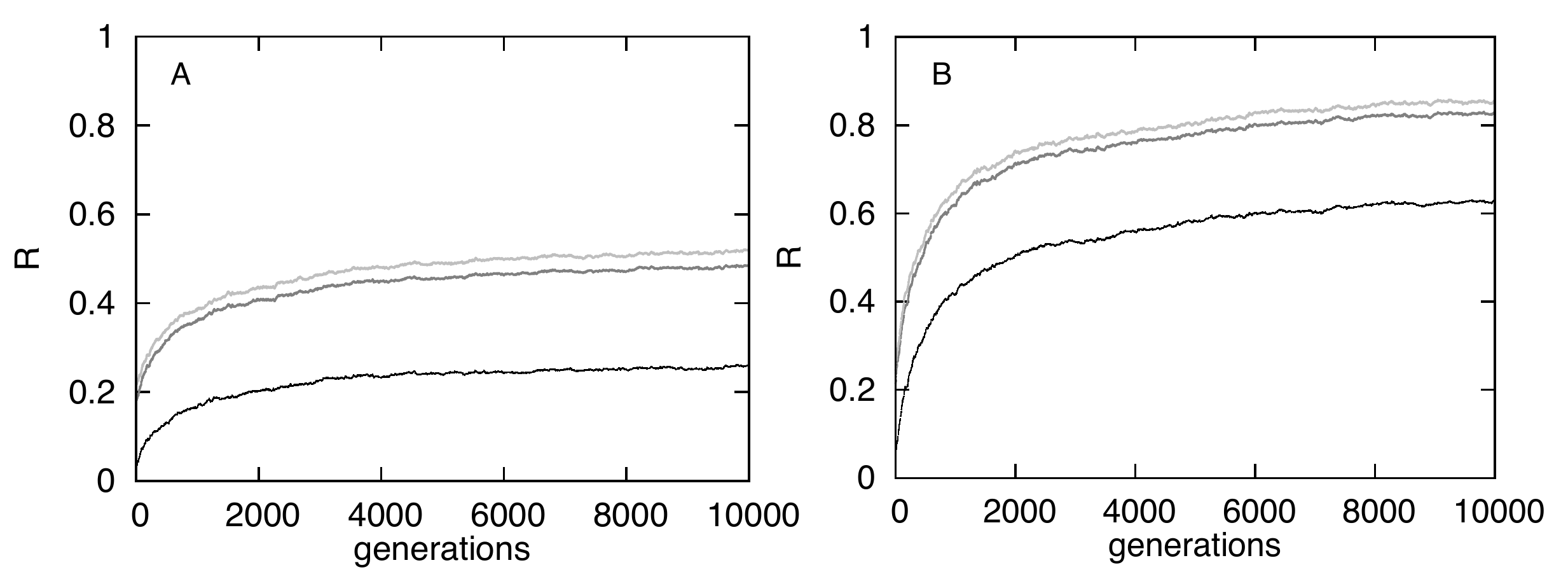}
\caption{Representation as a function of evolutionary time, for three different time intervals. A: Representation in ANNs (red: $R_t$, green: $R_{t-1}$, blue: $R_{t-2}$. B: Representation in Markov brains (colors as in A). The red curves are the same as in Fig.~\ref{fig:fit-rep}B, and are shown here for the purpose of comparison. }
\label{fig:R_memory}
\end{figure}

One can also ask whether brain states represent the environment as it is at the time it is being represented, or whether it represents the environment in its past state, or in other words, we can ask whether representations are about more distant or more proximal events. To answer this, we define {\em temporal representations} by including the temporal index of the Markov variables. For example, a representation at the same time point $t$ is defined [as implicit in Eq.~(\ref{rep})] as
\be 
R_t=H(E_t:M_t|S_t) \label{rept}\;,
\ee
while a representation of events one update prior is defined as
\be
R_{t-1}=H(E_{t-1}:M_t|S_t)\;,
\ee
\textit{i.e.}, the shared entropy between the internal variables at time $t$ and the environmental states at time $t-1$, given the sensor's states at time $t$. Naturally, one can define temporal representations about more distant events in the same manner. $R_t,R_{t-1}$, and $R_{t-2}$ were calculated and averaged over all 80 experiments in each generation (for both ANNs and MBs) over the course of evolution. Figure~\ref{fig:R_memory} shows that in both systems $R_{t-1}$ is larger than $R_t$, and $R_{t-2}$ is larger still (but note that $R_{t-3}$ is smaller, data not shown), and all values increase over evolutionary time similar to $R$ (see Fig.~\ref{fig:fit-rep}B). This suggest that both networks evolve a form of memory that reaches further back than just one update.

We suggest that the peak in representation at a time difference of two updates implied by Fig.~\ref{fig:R_memory} can be explained by the hierarchical structure of the networks, which have to process the sensorial information through at least two time steps to reach a decision (it takes at least two time steps in order to assess the direction of motion of the block). Decisions have to be made shortly thereafter, however, in order to move the agent to the correct location in time. This further strengthens our view that representations are evolved, and furthermore that they {\em build up during an agent's lifetime} as memory of past events shape the agent's decisions.

\section{Conclusions}

We defined a quantitative measure of representation $R$ in terms of information theory, as the shared entropy between the states of the environment and internal ``brain" states, given the states of the sensors. While internal states are necessary in order to encode internal models, not all internal states are representations. Indeed, representational information (which is about the environment) is a subset of the information stored in internal states. Information about the state of other past internal states or sub-states do not count towards $R$, and neither would information about imagined worlds, for example. Testing which nodes of the system contain representations about what aspect of the environment helps us to distinguish between information present in internal states and information that is specifically used as a representation. We applied this measure to two types of networks that were evolved to control a simulated agent in an active categorical peception task. Our experiments showed that the achieved $R$ increases with fitness during evolution independently of the system used. We also showed that while the (algorithmic) function of both artificial neural networks and Markov networks is difficult to understand, deterministic Markov networks can be reduced to sets of Boolean logic functions. This logic, however, may be epistemically opaque. While representation $R$ increases in networks over evolutionary time, each neuron can represent parts of individual concepts (features of the environment). However, most often concepts are distributed over several neurons and represent synergistically. In addition, representations also form over the lifetime of the agent, increasing as the agent integrates information about the different concepts to reach a decision. Thus, what evolves in Markov and artificial neural networks via Darwinian processes are not the representations themselves, but rather what evolves is the {\em capacity to represent the environment}, while the representations themselves are formed as the agent observes and interacts with the environment. We argued that $R$ can be measured in continuous (artificial neural networks) as well as discrete systems (Markov networks), which suggests that this measure can be used in more complex and more natural systems. We found that in the implementation used here, Markov networks were able to evolve their ability to form representations more easily than artificial neural networks. Future investigations will show what kind of system is more powerful to make intelligent decisions using representations.

\noindent{\bf Acknowledgements}
We thank C. Koch and G. Tononi for extensive discussions about representations, information integration, and qualia. This research was supported in part by the German Federal Ministry of Education and Research, by the Paul G. Allen Family Foundation, by the National Science Foundation's Frontiers in Integrative Biological Research grant FIBR-0527023, NSF's BEACON Center for the Study of Evolution in Action under contract No. DBI-0939454, as well as by the Agriculture and Food Research Initiative Competitive Grant no. 2010-65205-20361 from the USDA National Institute of Food and Agriculture. We wish to acknowledge the support of the Berlin-Brandenburg Academy of Sciences, the Michigan State University High Performance Computing Center and the Institute for Cyber Enabled Research. 
\bibliographystyle{apalike}
\bibliography{R.bib,Phi.bib}

\end{document}